\documentclass[12pt]{article}

\usepackage{amsmath,amssymb, fullpage}
\usepackage{graphicx}
\usepackage[latin1]{inputenc}
\usepackage[T1]{fontenc}
\usepackage{lmodern}
\usepackage[pdfstartview={FitH}]{hyperref}
\usepackage{algorithm}
\usepackage{algorithmic}
\newcommand{\lecture}[4]{
   \pagestyle{myheadings}
   \thispagestyle{plain}
   \newpage
   \setcounter{page}{1}
   \noindent
}
\newtheorem{theorem}{Theorem}
\newtheorem{lemma}{Lemma}

\newtheorem{defn}{Definition}

\newtheorem{notation}{Notation}

\def\beq{\begin{eqnarray}}
\def\eeq{\end{eqnarray}}
\def\beqs{\begin{eqnarray*}}
\def\eeqs{\end{eqnarray*}}

\newcommand{\Z}{\mathbb{Z}}

\newcommand{\R}{\mathbb{R}}

\newcommand{\con}{\mathbf{C}}
\newcommand{\aff}{\mathcal{A}}
\newcommand{\CC}{\mathcal{C}}

\newcommand{\A}{\mathbb{A}}

\newcommand{\E}{\mathbb{E}}
\newcommand{\p}{\mathbb{P}}

\newcommand{\one}{\mathbf{1}}

\newcommand{\ra}{\rightarrow}

\def\l{{\mathbf{\lambda}}}

\def\A{{\mathcal{A}}}
\def\ie{i.\,e.\,}

\def\var{{\mathrm{var}}}

\title{Distributed averaging in the presence of a sparse cut}
\author{Hariharan Narayanan
\\ Department of Computer Science,\\ University of Chicago
}

\begin{document}
\maketitle
\begin{abstract}
We consider the question of averaging on a graph that has one sparse
cut separating two subgraphs that are internally well connected.
 While
there has been a large body of work devoted to algorithms for
distributed averaging, nearly all algorithms involve only {\it
convex} updates. In this paper, we suggest that {\it non-convex}
updates can lead to significant improvements.  We do so by
exhibiting a decentralized algorithm for graphs with one sparse cut
that uses non-convex averages and has an averaging time that can be
significantly smaller than the averaging time of known distributed
algorithms, such as those of \cite{tsitsiklis, Boyd}. We use
stochastic dominance to prove this result in a way that may be of
independent interest.
\end{abstract}

\section{Introduction}
Consider a Graph $G = (V, E)$, where i.i.d
Poisson clocks with rate $1$ are associated with each
edge\footnote{This model can be simulated using previous models such
as \cite{Boyd} by allocating edges to nodes and equipping nodes with
multiple i.i.d poisson clocks.}. We
represent the ``true" real valued time by $T$. 
Each node $v_i$ holds a value $x_i(T)$ at time $T$. Let the average
value held by the nodes be $x_{av}.$ Every time an edge $e = (v, w)$
ticks, it updates the values of vertices adjacent to it on the basis
of present and past values of $v, w$ and their immediate neighbors
according to some algorithm $\A$.
 There is an extensive body of
work surrounding the subject of gossip algorithms in various
contexts. 
Non-convex updates have been used in
the context of a second order diffusion for load balancing
\cite{muthu} in a  slightly different setting. The idea there was to
take into account the value of the nodes during the previous two
time steps rather than just the previous one, (in a synchronous
setting), and set the future value of a node to a non-convex linear
combination of the past values of some of its neighbors.
 There is also a line
of research on averaging algorithms having two time scales,
\cite{borkar, konda} which is closely related to the present paper.

 In a previous
paper \cite{me}, we considered the use of non-convex combinations
for gossip on a geographic random graph on $n$ nodes. There we
showed that one can achieve averaging using $n^{1+ o(1)}$ updates if
one is willing to allow a certain amount of centralized control.
The main
technical difficulty in using non-convex updates is that they can
skew the values held by nodes in the short term. We show that
nonetheless, in the long term this leads to faster averaging.  Let
the values held by the nodes by $X(T) = (x_1(T), \dots,
x_{|V|}(T))^T$.
 We study distributed averaging algorithms $\A$ which
result in
$$\lim_{T \ra \infty} X(T) = x_{av} \one,$$ where $x_{av}$ is invariant under the passage of time.
 and show that in some
cases there is an exponential speed-up in $n$ if one allows the use
of non-convex updates, as opposed to only convex ones.

\begin{defn}
Let $$\var \, X(t) := \frac{\sum_{i=1}^{|V|} (x_i(t) -
x_{av})^2}{|V|}.$$ Let
$$T_{av} = \sup_{x \in \R^{|V|}}\inf_t   \,\,\, \p\left[\,\, \exists T >
t,\,\, \frac{\var \, X(T)}{\var \, X(0)}
> \frac{1}{e^2} \, \,\Big | \, \, X(0) = x\right] < \frac{1}{e}.$$
\end{defn}

\begin{notation}\label{not:fram} Let a connected graph $G = (V, E)$ have a partition into
connected graphs $G_1 = (V_1, E_1)$, and $G_2 = (V_2, E_2)$.
Specifically, every vertex in $V$ is either in $V_1$ or $V_2$, and
every edge in $E$ belongs to either $E_1$ or to $E_2$, or to the set
of edges $E_{12}$ that have one endpoint in $V_1$ and one in $V_2$.
Let $|V_1| = n_1$, $|V_2| = n_2$ where without loss of generality,
$n_1 \leq n_2$ and $|V| = n$. Let $T_{van}(G_1)$ and $T_{van}(G_2)$
be the averaging times of the ``vanilla" algorithm  that replaces at
the clock tick of an edge $e$ the values of the endpoints of $e$ by
the arithmetic mean of the two, applied to $G_1$ and $G_2$
respectively.
\end{notation}
\begin{defn} Let $\con$ denote the set of algorithms that use only convex
updates of the form
\begin{enumerate}
\item $x_i(t^+) = \alpha x_i(t^-) + \beta x_j(t^-)$.
\item $x_j(t^+) = \alpha x_j(t^-) + \beta x_i(t^-).$
\end{enumerate}
where $\alpha \in [0, 1]$ and $\alpha + \beta =1$.
\end{defn}
 These updates have been
extensively studied, see for example \cite{tsitsiklis, Boyd}.
\begin{theorem} The averaging time of any distributed algorithm in $\con$
is $\Omega(\frac{\min(|V_1|, |V_2|)}{|E_{12}|})$
\end{theorem}
\begin{theorem}
The averaging time of  $\aff$ is $O(\log n (T_{van}(G_1) +
T_{van}(G_2))).$
\end{theorem}
Note that in the case where $G_1$ and $G_2$ are sufficiently well
connected internally but poorly connected to each other, $\aff$
outperforms any algorithm in $\con$. In fact  for the graph $G'$
obtained by joining two complete graphs $G_1'$, $G_2'$ each having
$\frac{n}{2}$ vertices by a single edge, $\Omega(\frac{\min(|V_1'|,
|V_2'|)}{|E_{12}'|}) = \Omega(n)$, while $O(\log n (T_{av}(G_1') +
T_{av}(G_2'))) = O(\log n).$

\subsubsection{Algorithm $\aff$}
Let the vertices of $G_1$ be labeled by $[n_1]$ and those of $G_2$
by $[n_2]\setminus[n_1],$ where $[n] := \{1, \dots, n\}$. Let $e_c =
(v_{n_1}, v_{n_1+1}$ be a fixed edge belonging to $E_{12}$.
 Let the time of
the $k^{th}$ clock tick of an edge $e$ be $t$.  Let $C>>1$ be a
sufficiently large absolute constant (independent of $n$.)

\begin{algorithm}\label{alg:aff}
\begin{itemize}
\item If the edge $e$ is $e_c = (v_{n_1}, v_{n_1+1})$,
\begin{enumerate}

\item If $k \equiv -1 \,\,\, \mathrm{mod}\,\,\,  (\lceil C( T_{van}(G_1) + T_{van}(G_2))\ln n \rceil)$
\begin{enumerate}
\item $x_{n_1}(t^+) = x_{n_1}(t^-) + n_1\left\{x_{n_1+1}(t^-) - x_{n_1}(t^-)\right\}  $
\item $x_{n_1+1}(t^+) = x_{n_1+1}(t^-) - n_1\left\{x_{n_1+1}(t^-) - x_{n_1}(t^-)\right\}  $
\end{enumerate}
\item If $k \not\equiv -1 \,\,\, \mathrm{mod}\,\,\, (\lceil C( T_{van}(G_1) + T_{van}(G_2)) \ln n \rceil
)$ make no update.
\end{enumerate}
\item If the edge $e$ is $(v_i, v_j)  \not \in E_{12}$
\begin{enumerate}
\item $x_i(t^+) = \frac{x_i(t^-) + x_j(t^-)}{2}$.
\item $x_j(t^+) = \frac{x_i(t^-) + x_j(t^-)}{2}.$
\end{enumerate}
\item If $e \in E_{12} \setminus \{e_c\}$ make no update.
\end{itemize}
\end{algorithm}

\section{Limitations of convex combinations}

 Given a function $a(t)$, let  its right limit at $t$ be denoted by
$a(t+)$ and its left limit at $t$ by $a(t^-)$. Consider an algorithm $\CC \in \con$.\\
Let us consider the initial condition where $X(0)$ is the vector
that is $1$ on vertices $v_1, \dots, v_{n_1}$ of $G_1$ and
$-\frac{n_1}{n_2}$ on vertices $v_{n_1+1}, \dots, v_{n}$ of $G_2$.
Let us denote $\frac{\sum_{i=1}^{n_1} x_i(t)}{n_1}$ by $y(t)$ and
$\frac{\sum_{i=n_1+1}^{n_2} x_i(t)}{n_2}$ by $z(t)$. In the model we
have considered, with probability $1$, at no time does more than one
clock tick. \\In the course of the execution any algorithm in $\con$
$y(t)$ can change only during clock ticks of $e_c$ and the same
holds for $z(t)$. This is because during a clock tick of any other
edge, both of whose end-vertices lie in $G_1$ or in $G_2$, $y(t)$
and $z(t)$ do not change. The vertices adjacent to $e_c$ can change
by at most $2$ across these instants. Further, the values $x_n(t)$
and $x_{n+1}(t)$ are seen to lie in the interval
$$[\min_{i \in |V|} x_i(0), \max_{i \in |V|} x_i(0)] \subseteq [-1, 1].$$
If the clock of $e_c$ ticks at time $t$, we therefore find that
 \beq \label{e:1} |y(t^+) - y(t^-)| \leq \frac{2}{n_1},
\eeq The number of clocks ticks of $e_c$ until time $t$ is a Poisson
random variable whose mean is $t$.

A direct calculation tells us that \beq \var(X(t)) & \geq &
\frac{n_1 y(t)^2}{n}  .\eeq

To obtain a lower bound for ${y(t)}^2$, we note that the total
number of times the clocks of  edges belonging to $ E_{12}$ tick is
a Poisson random variable $\nu_t$ with mean $t|E_{12}|$. It follows
from Inequality~(\ref{e:1})  that $y(t) \geq 1 -
\frac{2\nu_t}{n_1}$.

\beqs |E_{12}| T_{av} & = & \E[\nu_{T_{av}}]\\
& \geq & \p\left[\nu_{T_{av}} \geq (1-\frac{1}{e})
\frac{n_1}{4}\right] (1 -\frac{1}{e})\frac{n_1}{4}\eeqs However
$$\p\left[\nu_{T_{av}} \geq (1-\frac{1}{e})
\frac{n_1}{4}\right]$$ must be large, because otherwise $y(T_{av})$
would probably be large. More precisely, \beqs  \p\left[\nu_{T_{av}}
\geq (1-\frac{1}{e}) \frac{n_1}{4}\right] & \geq & 1- \p\left[\,\,
\exists T
> T_{av},\,\, \var \, X(T)
> \frac{1}{e^2} \, \right]\\
&\geq & 1 - \frac{1}{e} \eeqs

Therefore, \beqs  T_{av} & \geq & \p\left[\nu_{T_{av}} \geq
(1-\frac{1}{e}) \frac{n_1}{4|E_{12}|}\right] (1
-\frac{1}{e})\frac{n_1}{4}\\
\geq  \Omega (\frac{n_1}{|E_{12}|})\eeqs


\section{Using non-convex combinations}

\subsubsection{Analysis}
 Since $T_{av}$ is defined in terms of variance and
algorithm $\aff$ uses only linear updates, we may subtract out the
mean from each $X_i(0)$ and it is sufficient to analyze the case
when $x_{av} = 0$. \\
Let $V_1 = [n_1]$ and $V_2 = [n]\setminus[n_1]$.  Let $\mu_1(t) =
\frac{\sum_{i=1}^{n_1} x_i(t)}{n_1}$ and $\mu_2 =
\frac{\sum_{i=n_1}^{2n} x_i(t)}{n}$ and $\mu(t) = |\mu_1(t)| +
|\mu_2(t)|.$ Let
$$\sigma(t) = \sqrt{\frac{\sum_{i=1}^{n_1} (x_i(t) - \mu_1(t))^2 +
\sum_{n_1 +1}^{n} (x_i(t) - \mu_2(t))^2}{n}}.$$ \\We consider time
instants $T_1, T_2, \dots$ where $T_i$ is the instant at which the
clock of edge $e$ ticks for the $\lceil iC (T_{van}(G_1) +
T_{van}(G_2)) \ln n \rceil^{th}$ time. Observe that the value of
$\mu(t)$ changes only across time instants $T_{k}, k = 1,
2, \dots$.\\
The amount by which $x_{n_1}(t)$ and $x_{n_1 + 1}(t)$ deviate from
$\mu_1(t)$ and $\mu_2(t)$ respectively, can be seen to be bounded
above by $\sqrt{n} \sigma(t)$

\beq \label{bridge} \max\left\{|x_{n_1}(t) - \mu_1(t)|\,\,,
|x_{n+1}(t) - \mu_2(t)|\right\} & \leq & \sqrt{n} \sigma(t).\eeq

 We now examine the
evolution of $\sigma(T_k^+)$ and $\mu(T_k^+)$ as $k \ra \infty$. The
statements below are true if $C$ is a sufficiently large universal
constant (independent of $n$).

From $T_{k}^+$ to $T_{k+1}^-$, independent of $x$, \beq \label{f:1}
\p\left[\sigma(T_{k+1}^-) \geq \frac{\sigma(T_{k}^+)} {n^6} \, \Big
|\, X(T_{k}^+) = x\right]
& \leq & \frac{1}{4n}\\
\mu(T_{k+1}^-) & = & \mu(T_k^+). \eeq

Because of inequality~(\ref{bridge}), from $T_{k}^+$ to $T_{k+1}^-$
\beq
\sigma(T_{k+1}^+) \leq n(\sigma(T_{k+1}^-) + |\mu(T_{k+1}^-)|)\\
|\mu(T_{k+1}^+)| \leq  n^{\frac{3}{2}} \,\,\sigma(T_{k+1}^-)\eeq
$$\var \,\, X(t) = \mu(t)^2 + \sigma(t)^2.$$
We deduce from the above that \beq \label{f:2}\p\left[\var \,\,
X(T_{k+1}^+) \geq \frac{\var \,\, X(T_k^+)}{n^4}\right] & \leq &
\frac{1}{4n}\eeq

Let $A_k$ be the (random) operator obtained by
composing the linear updates from time $T_k^+ $ to $T_{k+1}^+$. Let
$\|A\|$ denote the norm of an operator acting from  $\ell_2$ to
$\ell_2$
$$\|A\| = \frac{\sup_{x \in \R^{n}} \|Ax\|_2}{\|x\|_2}.$$

\begin{lemma}\label{l:1}
\beq \label{f:1} \p\left[\|A_k\|^2 \geq \frac{1}{n^3}\right] & \leq
& \frac{1}{2}\eeq
\end{lemma}
 To see this,
let $v_1, \dots, v_{n}$ be the canonical basis for $\R^{n}.$ For any
unit vector $$x = \sum_{i=1}^{n} \l_i v_i $$ Then, \beq
\|A_k(x)\| & \leq & \sum_{i=1}^{n} |\l_i| \,\,\|A_k(v_i)\|\,\,\,\,\,(\text{Triangle Inequality})\\
& \leq & \sqrt{\sum_{i=1}^{n}
\|A_k(v_i)\|^2}\,\,\,\,\,(\text{Cauchy-Schwartz inequality}) \eeq
The Lemma now follows from Inequality~(\ref{f:2}) by an application
of the Union Bound. {\hfill $\Box$} \\Moreover, we observe by
construction that the norm of $A_k$ is less or equal to $n$, \beq
\label{h:2} \|A_k\| & \leq & n  \eeq Note that $\log (\var \,
X(T_k^+))$ defines a random process (that is {\it not}  Markov). The
updates $A_k$ from time $T_k^+$ to $T_{k+1}^+$ for successive $k$
are i.i.d random operators acting on $\R^{2n}$. Note that $$\log
(\var \, X(T_k^+)) -\log (\var \, X(0)) \leq \sum_{i=1}^k \log
\|A_i\|$$ due to the presence of the supremum in the definition of
operator norm.
$$W_k := \sum_{i=1}^k \log \|A_i\|$$ is a random walk on the real
line for $k = 1, \dots, \infty$.

The last  and perhaps  most important ingredient is that of {\it
stochastic dominance}. It follows from Lemma~\ref{l:1} and
Equation~\ref{h:2} that the random walk $\{W_k\}$ can be coupled
with a random walk $\{\tilde{W}_k\}$ that is always to the right of
it on the real line, \ie for all $k$, $W_k \leq \tilde{W}_k$, where
the increments \beq \tilde{W}_{k+1} - \tilde{W}_k& = &\log n
\,\,\,\, (\text{with probability } \frac{1}{2})\\
& = & - \frac{3}{2} \log n \,\,\,\, (\text{with probability }
\frac{1}{2}.) \eeq

Noting that by construction, \beq \log (\var \, X(T_k^+)) - \log
(\var \, X(0)) & \leq & \tilde{W}_k ,\eeq it follows  that $T_{av}$
is upper bounded by any $t_0$ which satisfies
$$ \p\left[\,\, \forall T
> t_0,\,\, \tilde{W}_T \leq -2 \right] > 1 - \frac{1}{e}.$$

Note that $\E [\tilde{W}_k] = - \frac{k \log n}{2}$ and $\E [ \var
\tilde{W}_k] = \frac{9 k}{16} \log^2 n$.

 In order to proceed, we
shall need the following  inequality about simple unbiased random
walk $\{S_k\}_{k \geq 0}$ on $\Z$ starting at $0$.
\begin{theorem}
There exist constants $c, \beta$ such that for any $n \in \Z$, $ s >
0$
$$\p[S_n \geq s \sqrt{n}] \leq c e^{- \beta s^2}.$$
\end{theorem}

Using this fact, \beq \p[ \forall T
> t_0,\,\, \tilde{W}_T \leq -2] & = & \p[\forall T > t_0, (\log n)(S_T -
\frac{T}{2}) \leq -2]\eeq

For large $n$, this is the same as $$ \p[\forall T > t_0, S_T  <
\frac{T}{2}]  \geq   1 - \sum_{T>t_0} c e^{-\beta T/4}.$$ Clearly,
there is a constant $t_0$ independent of $n$ such that $ 1 -
\sum_{T>t_0} c e^{-\beta T/4} > 1 - \frac{1}{e}$. This completes the
proof.
 {\hfill $\Box$}

%
\section{Acknowledgement}
I am grateful to Dimitris Achlioptas, Vivek Borkar, Stephen Boyd and
Steven Lalley  for many helpful discussions.

\end{document}